**Capping Layer Effects on $Sb_2S_3$-based Reconfigurable Photonic Devices**

Ting Yu Teo[1,2]*, Nanxi Li[2]*, Landobasa Y. M. Tobing[2], Amy S. K. Tong[2], Doris K. T. Ng[2], Zhihao Ren[3], Chengkuo Lee[3], Lennon Y. T. Lee[2], and Robert Edward Simpson[4]*

[1] Singapore University of Technology and Design, 8 Somapah Road, Singapore 487372, Singapore

[2] Institute of Microelectronics (IME), Agency for Science, Technology and Research (A*STAR), 2 Fusionopolis Way, Innovis #08-02, Singapore 138634, Singapore

[3] Center for Intelligent Sensors and MEMS, Department of Electrical and Computer Engineering, National University of Singapore, 4 Engineering Drive 3, Singapore 117583, Singapore

[4] School of Engineering, University of Birmingham, Edgbaston, Birmingham B15 2TT, United Kingdom

Corresponding authors: tingyu_teo@mymail.sutd.edu.sg, linx1@ime.a-star.edu.sg, r.e.simpson.1@bham.ac.uk

**Abstract**

Capping layers are essential for protecting phase change materials (PCMs) used in non-volatile photonics technologies. This work demonstrates how $(ZnS)_{0.8}$-$(SiO_2)_{0.2}$ caps radically influence the performance of $Sb_2S_3$ and Ag-doped $Sb_2S_3$ integrated photonic devices. We found that at least 30 nm of capping material is necessary to protect the material from Sulfur loss. However, adding this cap affects the crystallization temperatures of the two PCMs in different ways. The crystallization temperature of $Sb_2S_3$ and Ag-doped $Sb_2S_3$ increased and decreased respectively, which is attributed to interfacial energy differences. Capped and uncapped Ag-doped $Sb_2S_3$ microring resonator (MRR) devices were fabricated and measured to understand how the cap affects the device performance. Surprisingly, the resonant frequency of the MRR exhibited a larger red-shift upon crystallization for the capped PCMs. This effect was due to the cap increasing the modal overlap with the PCM layer. Caps can, therefore, be used to provide a greater optical phase shift per unit length, thus reducing the overall footprint of these programmable devices. Overall, we conclude that caps on PCMs are not just useful for stabilizing the PCM layer, but can also be used to tune the PCM crystallization temperature and reduce device footprint. Moreover, the capping layer can be exploited to enhance light-matter interactions with the PCM element.

**1) Introduction**

Programmable photonic devices are essential for realizing many next-generation photonic integrated circuit (PIC) technologies. Current PIC technologies are mostly application-specific and consist of repetitive simple circuitry designs [1, 2]. Moreover, developing them involves a slow and costly iterative process, which hinders photonic innovations [1]. Having programmable PIC components allows sophisticated circuit designs to be developed efficiently. In the short term, more advanced and complex circuitry can be developed, which will inevitably generate a wider range of novel photonic devices and enhance application-specific systems like optical networks [3, 4] and sensors [5, 6]. In effect, programmable PIC chips will be analogous to electronic field-programmable gate arrays (FPGAs) [7]. We anticipate photonic FPGAs to become accessible to mainstream users, offering an inexpensive and efficient way to design and prototype new photonic systems.

Controlling amplitude and phase of light are important features in a programmable PIC. To this end, various tuning mechanisms such as thermo-optic [8-10], electro-optic [11, 12], and free-carrier dispersion effects [13-15] have been explored. However, these mechanisms are volatile, which requires a constant power supply to hold an optical state. In addition, the refractive index changes are small and implementing them requires large footprint [16]. With the lack of non-volatile tuning devices in current PIC platforms, photonic systems can become inefficient and have limited device scalability. For instance, in application-specific Optical Neural Networks, each electro-optic/thermo-optic tuned weighting component requires a 10 mW power supply to sustain an optical state [17]. Moreover, for photonic FPGAs to have comparable performance to their electrical counterparts, the programmable logic unit must also consist of non-volatile memory. Hence developing non-volatile programmable photonic devices is one of the key design criteria for next-generation PIC technologies.

Phase change materials (PCMs) have drawn interest from the photonics community in recent years due to their non-volatile switching characteristics [18, 19]. Incorporating PCMs into conventional photonic structures can bring us a step closer to programmable PIC technologies. PCMs can be easily integrated with PIC devices by depositing tens of nanometer thick PCM films above the waveguide using standard complementary metal-oxide-semiconductor (CMOS) processes [16]. The mode index of the waveguide, $n_{eff}$, can then be effectively tuned when the PCM is switched between the amorphous and crystalline structural states [20-22].

In recent years, PIC devices that are tuned by PCMs have been developed and extensively studied. These studies ranged from the design and fabrication of conventional photonic structures that incorporate conventional data storage PCMs [20, 22-25] to developing sophisticated PCM switching operations with on-chip heaters [26-29]. The plethora of advantages offered by PCMs over other tuning mechanisms are clear, however, the common data storage PCMs are lossy at visible and near-infrared frequencies. Recently, therefore, state-of-the-art PCM-tuned photonic devices using low-loss PCMs, like $Sb_2S_3$ and $Sb_2Se_3$, have been proposed as better PCM alternatives to the data storage PCM, $Ge_2Sb_2Te_5$ [30, 31]. The introduction of wide band gap PCMs has led to the rapid development of low-loss PCM based photonic devices [26, 27, 32-36].

As low-loss PCMs are becoming popular choices for developing tunable photonic devices, further understanding of their properties and limitations is required for better device integration. One important area of study involves the capping layer, which is needed to protect low-loss PCMs from the atmosphere. Whilst the influence of capping layers on well-known PCMs like $Ag_4In_3Sb_{67}Te_{26}$ [37] or those on the $GeTe-Sb_2Te_3$ pseudobinary tie-line [38-43] have been extensively studied, understanding how capping layers influence the material, crystallization and optical properties of low-loss PCMs is still nascent. As such, capping layers were mainly considered for PCM protection purposes, and a wide range of capping layer thicknesses were used to justify material loss prevention during the structural phase transition heating process [26, 31, 32, 35, 44, 45]. This ranged from 10 nm [26, 32] to the hundreds of nanometers [31, 45]. Inadequate capping thickness can affect photonic device performance [32]. Moreover, capping layers can affect the PCM crystallization temperature [37-42] and $n_{eff}$ of the photonic device. These changes will influence the PCM switching parameters and the photonic device design.

Herein, we describe the influence of $(ZnS)_{0.8}$-$(SiO_2)_{0.2}$ capping layers on the structural phase transition and optical properties of two related low-loss PCMs, $Sb_2S_3$ and Ag-doped $Sb_2S_3$. These properties can influence the device design and switching operations. We chose to focus on $Sb_2S_3$ as it exhibits the lowest optical loss amongst the four PCMs, $Ge_2Sb_2Te_5$, $Ge_2Sb_2Se_4Te_1$, $Sb_2Se_3$ and $Sb_2S_3$ [36], and consequently it is of particular interest for tuning a wide range of photonic devices [19, 32, 34-36, 44, 46, 47]. Previously [48, 49], we showed that lightly doping $Sb_2S_3$ with Ag can lower the crystallization temperature and produce smaller crystallites. The melting point for lightly Ag-doped $Sb_2S_3$ (below 12.5 %) along the $Sb_2S_3$ – $Ag_2S$ tie line is also lower [50], suggesting a lower amorphization temperature. For these reasons, we believe that Ag-doped $Sb_2S_3$ has potential applications in tunable photonic devices. $(ZnS)_{0.8}$-$(SiO_2)_{0.2}$, a capping material widely used in phase change rewriteable devices, was chosen to cap the low-loss PCMs. This material can withstand high temperature phase transition processes [31, 40, 41, 51, 52] and has negligible optical losses in the visible and near infrared spectrum [53].

The capping layer should not be neglected when designing and developing low-loss PCM devices. Through thermogravimetric analysis (TGA) and film thickness measurements, we observed material loss during the heating process. Capping thickness of more than 30 nm were necessary to preserve the material. The crystallization behavior of capped $Sb_2S_3$ and Ag-doped $Sb_2S_3$ were also altered. Cap-PCM interface stumped crystal growth and altered nucleation behavior, which in turn caused the crystallization temperature of $Sb_2S_3$ and Ag-doped $Sb_2S_3$ to increase and decrease, respectively. This finding suggests that the device switching parameters of differently capped PCMs will vary. By incorporating Ag-doped $Sb_2S_3$ into a microring resonator (MRR) device, we found that the capping layer caused a larger red-shift than its uncapped form. Counterintuitively, the optical losses were also greater despite the capping material being transparent to the near infrared light. The larger red-shift and higher optical loss were due to an upward modal shift from the waveguide to the PCM layer. The greater mode confinement on the PCM layer suggests that overall device footprints can be reduced, thus mitigating the large footprint issue reported in low-loss PCM devices due to their smaller $\Delta n_{eff}$ [31, 32, 36]. Future low-loss PCM device designs should optimize capping layer parameters to enhance both device and system performance.

## 2) Methods

### PCM TGA Measurements

A thermogravimetric analyzer (TA-Instruments Q-50) was used to conduct TGA on the three PCMs: $Sb_2S_3$, $(Ag)_{0.1}$-$(Sb_2S_3)_{0.9}$ and $Ge_2Sb_2Te_5$. PCM dust residuals were collected by scrapping the surface of sputtering targets of 99.99% purity. The dust was then transferred into $Al_2O_3$ crucibles for heating in the thermogravimetric analyzer furnace. The samples were heated to 600 °C with a heating rate of 5 °C/min. To prevent material oxidation, $N_2$ gas was supplied at 40 mL/min to the heating chamber.

### Thin Film Deposition and Characterization

Amorphous $Sb_2S_3$, $Ag_{2.4}Sb_{42.2}S_{55.4}$, and $(ZnS)_{0.8}$-$(SiO_2)_{0.2}$ (simplified as ZnS-SiO$_2$ in the remaining text) thin films were deposited on Si substrates using radio frequency confocal magnetron sputtering system, AJA Orion 5, at room temperature. All films were deposited from commercially available targets of 50.8 mm diameter and 99.99% purity. The $Ag_{2.4}Sb_{42.2}S_{55.4}$ (simplified as Ag-SbS in the remaining text) films came from a $(Ag)_{0.1}$-$(Sb_2S_3)_{0.9}$ target. We confirmed the thin film composition with scanning electron microscope-energy dispersive X-Ray (SEM-EDX). The EDX results can be found in the main supporting information document, Table S1. All depositions were done at a process pressure of 0.5 Pa, with a base pressure below $6.7 \times 10^{-5}$ Pa. The deposition power for $Sb_2S_3$, Ag-SbS, and ZnS-SiO$_2$ were 30 W, 40 W and

100 W respectively, and their corresponding deposition rates were 0.69 Å/s, 0.55 Å/s and 0.76 Å/s. The thicknesses of the films were confirmed with profilometry scans using an atomic force microscope (AFM), Asylum Research MFP-3D Origin.

To determine whether the film thickness changes during thermal annealing, step profilometry of films were also conducted with the AFM before and after crystallization. The AFM scan areas were 40 $\mu$m by 40 $\mu$m wide and contained 512 scanning lines. An example of the AFM scan and film thickness derivation can be found in the main supporting information document, Figure S1. The film thickness is represented as the average thickness across seven different regions of the wafer sample. $Sb_2S_3$ and Ag-SbS films of different capping thicknesses were crystallized at 320 °C and 280 °C respectively in a heating furnace (Linkam T95-HT), with a ramp rate of 5 °C/min and then held at their corresponding crystallization temperatures to ensure complete PCM crystallization. The hold time typically took between 20 to 30 min. To prevent film oxidation, the furnace chamber was purged with Ar gas at 4 sccm during the experiment.

To compare the crystallization temperature of capped and uncapped PCMs, the samples were also heated in the Linkam furnace at 5 °C/min from room temperature to 350 °C. Similarly, the chamber was purged with Ar gas at 4 sccm. The Linkam microscope furnace was attached to a microscope-camera setup and a spectrometer (Ocean Optics flame). The spectrometer was used to measure the spectrum of a light source (Olympus TH4 200), which was reflected from the thin film samples during heating. The light source wavelength ranged from 500 nm to 700 nm. In this experiment, a microscope image and reflectivity reading of the film were taken for every 1 °C rise in temperature. The spectra and images were then analyzed to find the non-isothermal crystallization behavior of the films.

Ellipsometry of films was conducted using the Accurion EP4 system to extract the material optical constants. Delta, $\Delta$, and Psi, $\Psi$, measurements were taken across the 400 nm to 900 nm spectral band. The data was then fitted with Tauc Lorentz and Lorentz oscillator models [36, 54]. The complex refractive indices ($n$+i$k$) for the $400 nm < \lambda < 1600 nm$ wavelength range were found by curve-fitting. The $n$ and $k$ values from 900 nm to 1600 nm were extrapolated based on the fitted model. These extrapolations were valid as previous ellipsometry measurements at near infrared wavelengths show the fitting models converging asymptotically towards a single $n$ and $k$ value after the material bandgap wavelength, which is located in the visible wavelength range [32, 34, 36, 53].

**Waveguide Mode Simulations**

Finite difference eigenmode (FDE) solver from Lumerical MODE [55] was used to solve Maxwell's equations for the electromagnetic wave distribution within the PCM-deposited waveguide model. Hence, the $n_{eff}$ and loss values of the waveguide structure can be obtained. In the simulation model, optical constants of the low-loss PCMs and cap were obtained from ellipsometry measurements described above, whilst $Si_3N_4$ and $SiO_2$ were obtained from literature [56, 57].

**Ag-SbS-tuned MRR Device Fabrication and Characterization**

MRR devices were fabricated by a Si photonic foundry. The devices were made of $Si_3N_4$ waveguides that were 1.5 $\mu$m wide, 200 nm thick, and were embedded in a 2 $\mu$m thick $SiO_2$ cladding. The top $SiO_2$ cladding at the coupling region was removed to expose the waveguide for PCM deposition. The removal process involved immersing the chips in 10 % HF solution for 3 min followed by a deionized water stop bath. Upon removing the top SiO$_2$ cladding, PCM patches (40 $\mu$m by 40 $\mu$m) were patterned on the devices using maskless photolithography (Nanyte BEAM). Ag-SbS and the corresponding ZnS-SiO$_2$ capping material were then deposited as described above, followed by a lift-off process which involved sonicating the chips at low power in acetone followed by isopropyl alcohol for 20 s each. The chips were then measured using a tunable laser source-photodetector system (Keysight 8164B), with the laser source being attached to a polarization controller (Thorlabs FPC 562). Afterwards, the Linkam furnace was used to crystallize the PCM patch for the crystalline PCM MRR device measurements. The annealing process for the capped and uncapped samples involved heating the furnace to 280 °C and 300 °C respectively at 5 °C/min and holding it for 30 min. The chamber was also purged with Ar gas at a flowrate of 4 sccm to prevent oxidation of the films.

**3) Results and Discussion**

The thermal stabilities of the two low-loss PCMs were found to be lower than Ge$_2$Sb$_2$Te$_5$, which highlights the importance of a capping layer in low-loss PCM tuned photonic devices. The TGA measurements in Figure 1 show that the mass loss for Sb$_2$S$_3$ and (Ag)$_{0.1}$-(Sb$_2$S$_3$)$_{0.9}$ occurred at a lower temperature than Ge$_2$Sb$_2$Te$_5$. We noticed an accumulated 4.5% mass loss in Sb$_2$S$_3$ between 150 °C to 270 °C followed by rapid mass loss starting from 500 °C, which is near the melting point of Sb$_2$S$_3$ [50]. A similar trend was also reported in previous literature [58]. The accumulated 4.5% mass loss between 150 °C to 270 °C was due to small amounts of Sulfur evaporating. The rapid mass loss at 500 °C was due to Sb$_2$S$_3$ melting and evaporating.

Doping the Sb$_2$S$_3$ compound with Ag helps to prevent Sulfur loss, thereby improving its thermal stability. We observed that the onset peak mass loss in (Ag)$_{0.1}$-(Sb$_2$S$_3$)$_{0.9}$ shifted from 260 °C to the higher temperature of 450 °C. Note, the mass loss was also less severe than the pure Sb$_2$S$_3$ due to the formation of AgSbS$_2$, Sb$_2$S$_3$ and Sb compounds upon doping [50]. The melting point of AgSbS$_2$ was found to be 450 °C, which accounts for this slight mass loss. Similarly, the compound experienced a rapid mass loss at temperatures close to the melting point of Sb$_2$S$_3$, indicating liquid Sb$_2$S$_3$ evaporating. The residual Sb did not contribute to the mass loss because the melting point of Sb is above 600 °C [50]. In the remaining studies, the Ag concentration was approximately 2 at.%, which is substantially lower than that of the TGA-studied sputtering target dust. We infer that the thermal stability of the thin film will be better than that of Sb$_2$S$_3$ but lower than the 10% Ag-doped target dust. Hence, adding a capping layer is important in preserving the Ag-SbS films.

Fine tuning the Ag concentration is important for optimizing Ag-SbS photonic devices. Introducing Ag into Sb$_2$S$_3$ helps to prevent Sulfur vaporization due to the formation of AgSbS$_2$, Sb$_2$S$_3$ and Sb compounds. Sulfur residuals seem to result in poor thermal stability, which is likely due to its low melting point [50]. Ag seems to help improve the thermal stability of Sb$_2$S$_3$, however, adding more than 8 at. % Ag will cause AgSbS$_2$ and Sb$_2$S$_3$ compounds to phase separate upon crystallisation [48, 49]. Moreover, Ag nanoparticles will emerge, which will absorb photons strongly and substantially reduce optical intensity through local plasmonic resonances [59, 60]. The 2% Ag-doped films, which were used in this work, do not show any evidence of Ag nanoparticulate precipitation. However, our future works will further optimize the Ag concentration to fine tune the crystallization kinetics and optical properties.

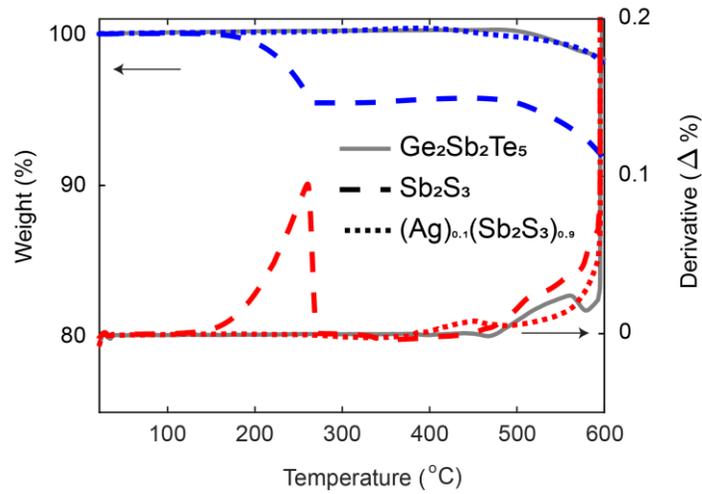

Figure 1: TGA measurements of $Sb_2S_3$, 10% Ag-doped $Sb_2S_3$ and $Ge_2Sb_2Te_5$.

With Sulfur loss occurring within the phase transition temperature range, the optimal capping layer thicknesses for the two low-loss PCMs were investigated. $Sb_2S_3$ and Ag-SbS films of various capping thickness were measured with AFM before and after crystallization. We associate the change in overall film thickness from the crystallization of PCM films [61, 62]. Note, ZnS-$SiO_2$ exhibits minimal thickness change upon thermal annealing [63], which was also verified in this work. The corresponding ZnS-$SiO_2$ thickness change measurements can be found in the main supporting information document, Table S2. We first studied the thermal stability of the $Sb_2S_3$ and Ag-SbS PCM layers, as shown in Figure 2a, which gives the PCM thickness reduction for different capping layer thicknesses upon thermal annealing. The thicknesses of the PCM materials were chosen as 33 nm for $Sb_2S_3$ and 30 nm for Ag-SbS because they are similar to the typical PCM thicknesses used to control photonic devices [21, 23, 26, 32, 34-36].

The PCM layer thickness shrunk by approximately 10% for the thickest caps. Indeed, the shrinking effect saturates for cap thicknesses of > 30 nm in Figure 2a. Thus, we conclude that at least 30 nm of ZnS-$SiO_2$ is needed to prevent $Sb_2S_3$ and Ag-SbS films from material degradation. This 10% thickness change represents an increase in the PCM density upon crystallization, which was also previously reported [61, 62]. A larger thickness change was seen for films with capping layers thinner than 30 nm. This larger change is likely due to cap fracturing, which allows the Sulfur atoms in both PCMs to escape. We observed distinct dark circular rings appearing at the nucleation sites of both uncapped crystalline films as shown in Figures 2b(i) and 2b(iv). After depositing a cap on the PCM films, the rings disappeared during annealing as can be seen in Figures 2b(iii) and 2b(vi). These rings could be signs of the PCM degrading. The uncapped $Sb_2S_3$ experienced a 30% thickness reduction, which was also reported in earlier works of uncapped films [30]. A lower thickness reduction was seen in the uncapped Ag-SbS films as Sulfur vaporization was mitigated. Based on these results, we recommend the use of at least 30 nm ZnS-$SiO_2$ to protect low-loss PCM-tuned photonic devices. For the rest of this work, the following capped samples were used in our analysis: (1) $Sb_2S_3$: 55 nm ZnS-$SiO_2$ cap on 33 nm $Sb_2S_3$ (2) Ag-SbS: 60 nm ZnS-$SiO_2$ cap on 30 nm Ag-SbS.

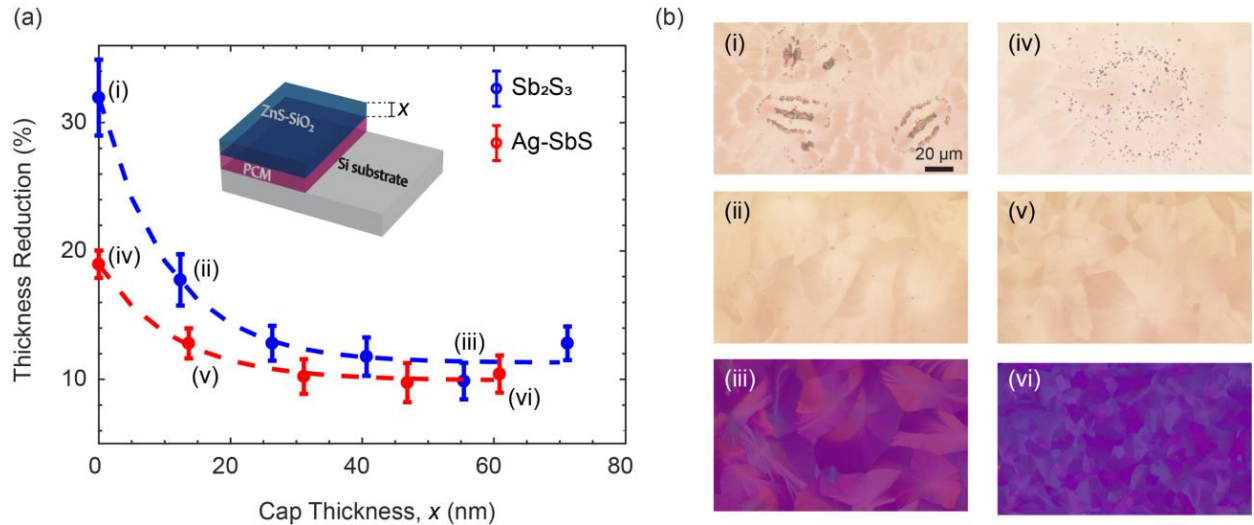

Figure 2: Capping layer thickness optimization. (a) Thickness reduction of both PCMs upon crystallization for different capping layer thicknesses. (b) Optical microscope images of crystalline PCM films with different cap thickness. (i) to (vi) show the corresponding cap thickness indicated in (a). Images were taken under the same lighting conditions and lens magnification.

The crystallization behaviors of capped and uncapped low-loss PCM films were investigated. Capped and uncapped $Sb_2S_3$ and Ag-SbS were deposited on Si substrates and heated under the same heating conditions (see methods section). We note that in previous works [37, 39-41], the samples used to investigate the capping layer influence were symmetrical, where the PCM was sandwiched between the capping materials. However, the films in this work were deposited directly on Si substrates because in programmable Si photonic devices, the PCM should ideally be in direct contact with the Si waveguide to maximize its interaction with the optical mode. The crystallization temperature is seen as an abrupt change in reflectivity [41, 64]. The changes in reflectivity of the capped and uncapped PCMs, and their corresponding microscope images at different temperatures during the annealing process, are presented in Figure 3. Whilst the reflectance of chalcogenide PCMs is known to increase upon crystallization [64, 65], we note that adding the $ZnS-SiO_2$ layer caused interference effects and the reflectivity readings of the capped samples decreased upon crystallization, as seen in Figures 3b and 3d. This trend was confirmed by the calculated reflectivity spectrum using the transfer matrix method on OpenFilters [66]. The reflectivity values of the amorphous capped samples were mainly higher than its crystalline form within the light source wavelength range. The simulated reflectivity spectrum of the four samples can be found in the main supporting information document, Figure S2.

The capping layer was found to influence the crystallization behavior of the low-loss PCMs. Based on the reflectivity measurements and microscope images in Figure 3, we make the following observations: (1) The reflectivity curve for capped $Sb_2S_3$ in Figure 3b experienced a more gradual change in reflectivity upon crystallization when compared to its uncapped form in Figure 3a. Conversely, the capped Ag-SbS in Figure 3d shows a sharper change in reflectivity when compared to its uncapped form in Figure 3c. (2) The presence of a cap impeded $Sb_2S_3$ and Ag-SbS crystal growth as can be seen in Figures 3e(iii) and 3f(iii).

The aforementioned differences in crystallization behavior can be attributed to the PCM-cap interfacial energy affecting the two crystallization mechanisms in a crystallization process: nucleation and growth [39, 41, 67]. Introducing an extra layer of material above the PCM changes the surface energy of the PCM by creating new chemical bonds and also mechanical stresses, which in turn affect the crystallization kinetics. The first observation shows the effect of capping layer on PCM nucleation. In capped $Sb_2S_3$, we observed a more gradual decrease in reflectivity during crystallization than its uncapped form. The gradual decrease was also observed in certain capped materials on $Ge_1Sb_2Te_4$ previously, which was caused by a larger surface energy difference between the PCM and cap [41]. Video 1 in the supporting information section shows the crystal matrix expanding outwards over a large temperature range. For the capped Ag-SbS, we observed a sharper decrease in reflectivity than when it is uncapped, which indicates the two mechanisms occurring concurrently. This result suggests that Ag-SbS has a similar interface energy to the $ZnS-SiO_2$ cap. Video 2 in the supporting information section shows the crystal nucleating and growing over a small temperature range. For the second observation, we compared the interfacial energies of PCM-cap with respect to the PCM structural states. The interfacial energy difference of crystalline PCM-amorphous cap is generally higher than amorphous PCM-amorphous cap [39]. As smaller interfacial energy differences are preferred, the cap prevents the $Sb_2S_3$ and Ag-SbS crystallites from radially growing, as shown in Figures 3e(iii) and 3f(iii), i.e. crystal growth of the two low-loss PCMs appears to be impeded by the capping layer. However, the films can still nucleate new seeds from which the crystal can grow. Consequentially, we observed a high density of small nuclei in the capped samples. We note that in this work, the capped $Sb_2S_3$ and Ag-SbS led to higher and lower crystallization temperatures respectively when compared to their uncapped forms. These results indicate that a cap can alter the PCM crystallization temperature and affect the switching energy used for structural phase transition. This effect is also important because $Sb_2S_3$ films are known to crystallize into large birefringent grains [47], but if the correct choice of cap can reduce the size of these grains such that they are substantially smaller than the wavelength of the optical signal, then polycrystalline films with an average isotropic refractive index tensor will result.

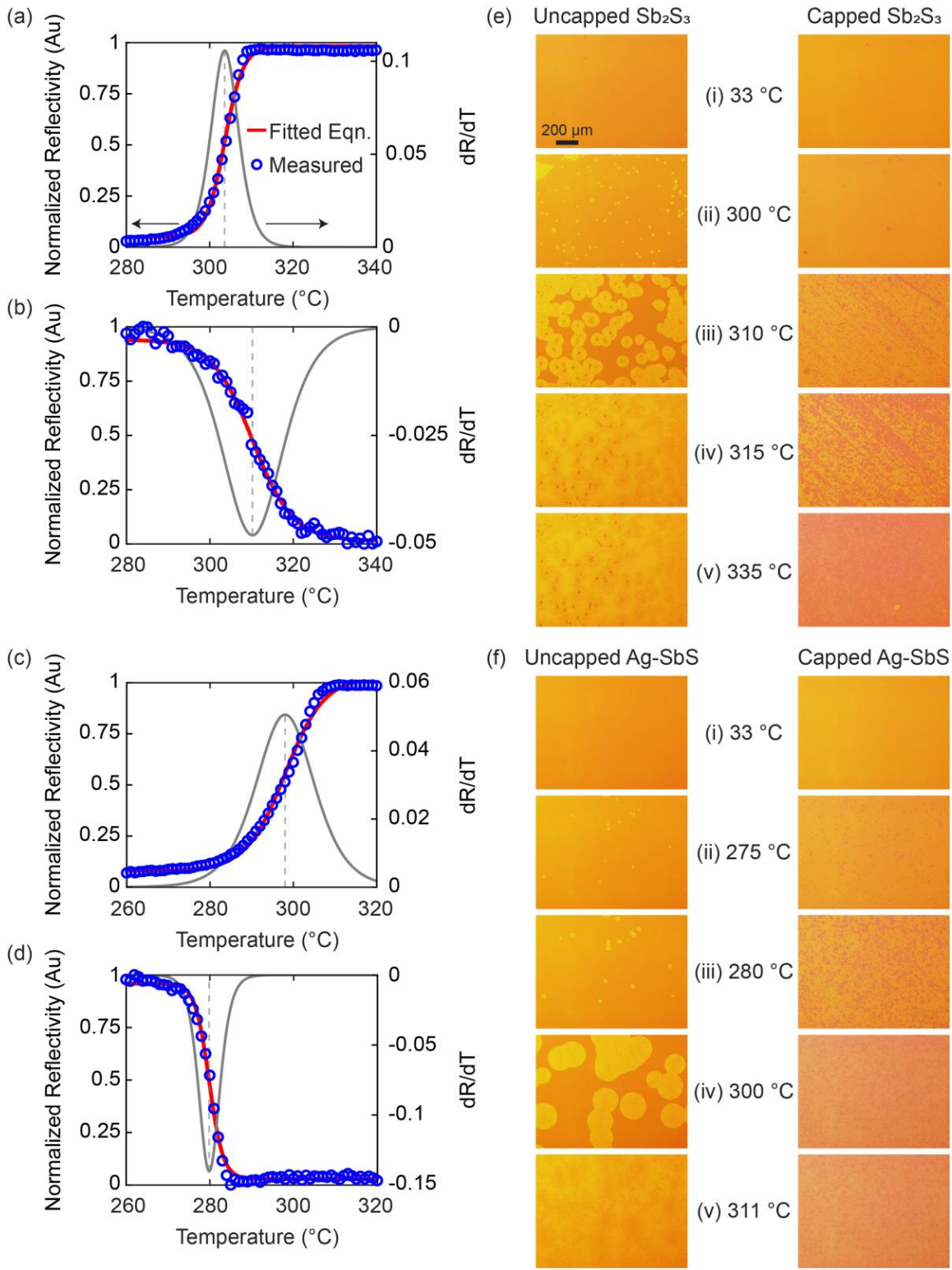

Figure 3: Crystallization profiles of capped and uncapped $Sb_2S_3$ and Ag-SbS. Reflectivity measurements of (a) uncapped and (b) capped $Sb_2S_3$, and (c) uncapped and (d) capped Ag-SbS with respect to temperature. Optical microscope images of (e) $Sb_2S_3$ and (f) Ag-SbS films during the crystallization process. (i) to (v) show the crystals nucleating and growing as the temperature increases.

To design PCM-tuned photonic devices, we first obtained the complex refractive index of the PCMs and their capping layers from ellipsometry measurements. Figure 4a shows the extracted ($n$, $k$) values for $Sb_2S_3$ and Ag-SbS PCMs in their crystalline and amorphous states, while Figure 4b shows the extracted ($n$, $k$) for the 70-nm thick $ZnS-SiO_2$ capping layer. The refractive index data shown in Figure 4 can be found in the second supporting information document (Supporting Information 2) for the interested readers. Extracting the optical constants of the crystalline PCMs with caps were non-trivial because a double-layer fitting model was required. Therefore, a single-layer $ZnS-SiO_2$ film was first measured and then incorporated into the double-layer film model. In this characterization, the single layer $ZnS-SiO_2$ film was also annealed at the 320 °C crystallization temperature to observe for any changes in the ($n$, $k$) values during annealing. As shown in Figure 4b, we found the ($n$, $k$) values to be consistent before and after annealing. Small changes in the refractive index were also observed in thicker films [63], which is consistent with our measurements.

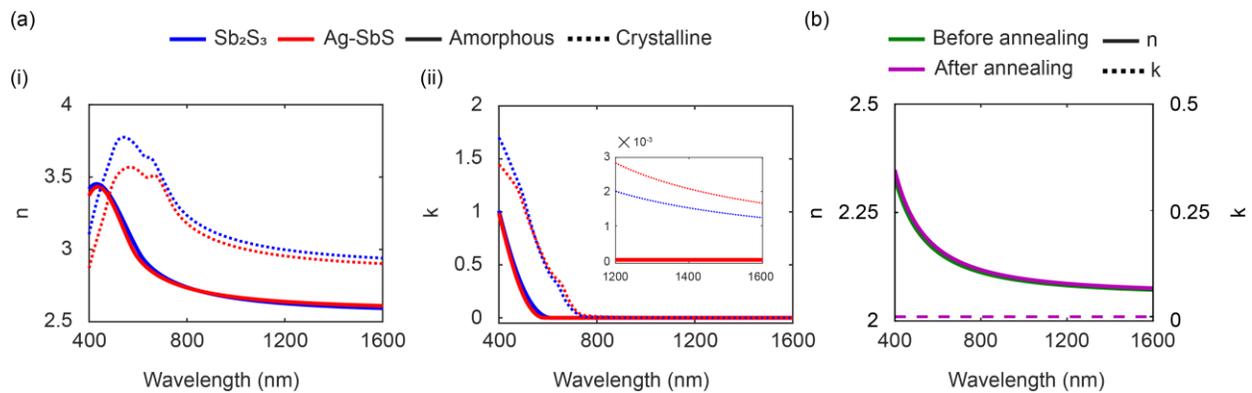

Figure 4: (a) Optical constants of low-loss PCMs $Sb_2S_3$ and Ag-SbS in the amorphous and crystalline states, with (i) and (ii) corresponding to the real and imaginary components respectively. Figure inset in (ii) shows the magnified $k$ data from 1200 nm to 1600 nm wavelength. (b) Optical constants of 70 nm $ZnS-SiO_2$ before and after annealing.

Owing to its higher thermal stability and lower crystallization temperature, we integrated Ag-SbS into $Si_3N_4$ MRR devices (125 μm radius). The $Si_3N_4$ devices were fabricated by a standard Si photonic foundry. As the pristine $Si_3N_4$ MRR devices were by default embedded in 2μm $SiO_2$ cladding, the cladding was first removed by HF wet etching to expose the waveguide for PCM deposition. However, due to fabrication limitations, the etching rate at the ring-bus coupling region was faster than the other regions, as shown in the Transmission Electron Microscopy (TEM) images in Figure 5a. For this reason, the PCM had to be deposited at the coupling region. Note, the $SiO_2$ top cladding outside the coupling region appeared uniform across the long bus waveguide. Hence the device outside the coupling region was protected by the unetched top $SiO_2$ cladding, and mode propagation was consistent throughout the device. The TEM cross section images of the etched device taken at different regions can be found in the main supporting information document, Figure S3.

With the $ZnS-SiO_2$ capping layer, the change in optical properties of the PCM layer upon crystallisation was amplified. This loading effect caused the optical mode to shift towards the PCM and experienced more spatial overlap with it. The $ZnS-SiO_2$ cap led to a higher $n_{eff}$ and, counterintuitively, a higher associated optical loss upon PCM crystallization. Figure 5b shows the Transverse Electric (TE) mode profile of the coupling region deposited with capped and uncapped Ag-SbS, together with their corresponding $n_{eff}$ and loss values. There was a higher mode spatial overlap with the PCM layer for the capped case as we see the mode pattern shifting towards the capping layer. This upward modal shift resulted in a higher $\Delta n_{eff}$ (by ~4x) and losses (by ~2x) as the PCM crystallizes. Note, the crystalline Ag-SbS layer undergoes thickness reduction in a similar manner to that shown in Figure 2b. Based on these findings, the capping layer provides an additional degree of freedom in the design of PCM-tuned photonic devices, whereby its thickness and optical properties can optimize $\Delta n_{eff}$ and loss to desired values.

Such a large change in $n_{eff}$ results in a red-shift in the MRR resonant wavelength. Moreover, the change in both $n_{eff}$ and loss led to different coupling conditions between the microring and the single waveguide channel. Coupling condition is dependent on the (1) mode spatial overlap between the two coupled waveguides and (2) amplitude cavity loss in the microring. These two parameters are denoted as the self-coupling coefficient, $r$, and cavity loss coefficient in the microring, $a$, respectively [68]. Indeed, the former is associated with the coupling strength of the paired waveguide and thus $n_{eff}$, whilst the latter is inversely related to the optical losses. The coupling condition depends on the matching between the coupling strength and cavity roundtrip loss. The system is in under-coupled situation when the roundtrip loss is greater than the amount of light coupled into the cavity, while the system is in over-coupled situation when the coupling strength is greater than the roundtrip loss. The system is in critically-coupled condition when the coupling strength is equal to the cavity loss ($r = a$), which corresponds to the highest transmission contrast due to optimized interference condition at the output port. With a change in both $r$ and $a$ in our devices, the coupling conditions changed from over-coupled to near critically-coupled or under-coupled upon crystallizing, as illustrated in the comparison of amorphous and crystalline transmission spectra of the various MRRs in Figure 5c.

To further study the effects of the cap on resonance shift, we tracked the MRR device azimuthal mode number [68], $m$, of the four devices, labelled as (i) to (iv) in Figure 5c. $m$ represents the integer multiple of wavelengths that fit the cavity length (ring circumference) at resonance wavelength. With an increase (decrease) in $n_{eff}$, the optical pathlength increases (decreases) and longer (shorter) wavelengths are required to interfere constructively during resonance under the same $m$ multiple. This causes the resonance peak to shift to a longer (shorter) wavelength. The $m$ value approximations of the devices can be found in the main supporting information document, Tables S3 to S6. The analysis focused on the resonant peaks near the 1550 nm wavelength.

The corresponding $m$ values of the resonance peak near the 1550 nm wavelength are shown in Figure 5c. The MRR device without the PCM layer corresponds to $m = 759$. Upon amorphous Ag-SbS deposition, the optical path length increased as a section of the ring has a higher $n_{eff}$. Hence for the same $m$ multiple, longer wavelengths were required to fit the entire ring circumference. For this reason, the resonance peak of $m = 759$ shifted to a longer wavelength as shown in Figure 5c(ii). Under the influence of a cap, the optical pathlength further increased and the shift becomes more pronounced. The amorphous $m$ value near the 1550 nm wavelength increased to $m = 761$ as shown in Figure 5c(iii). There was also a further increase in $m$ when the PCM layer thickness increased to 60 nm as shown in Figure 5c(iv). Within the device itself, we also observed a larger resonance peak shift for the capped PCM devices upon crystallization. With a cap on the PCM, the spectrum exhibited close to $2\pi$ phase shift (or close to ~FSR), as shown in Figures 5c(iii) and 5c(iv). By contrast, the uncapped PCM in Figure 5c(ii) showed a phase shift of less than $\pi$ (or less than 0.5 FSR). We note that the smaller phase shift could also be partly due to material degradation [32]. With a larger phase shift, the capped PCM length can be shortened to reduce the phase shift down to $\pi$. A shorter PCM length could mitigate the large footprint limitation reported in low-loss PCM photonic devices [31, 32, 36] as they have a smaller $\Delta n_{eff}$ when compared to the data storage PCM alloys along the GeTe-Sb$_2$Te$_3$ pseudobinary tie-line.

Next, we analysed the effects of optical loss on the waveguide coupling condition. Our analysis focused on the change in coupling condition within a single device upon structural phase transition as the optical losses across all devices in the amorphous state were negligible. As mentioned, the device coupling condition is determined by the matching between $r$ and $a$ [68]. $r$ is related to the coupling strength of the ring-bus waveguide, which corresponds to the difference between the even and odd modes of the coupled waveguide pair. $a$ is inversely related to the optical loss. For devices 5c(ii) and 5c(iii), $r$ was approximated to be consistent in both PCM structural states, whilst for device 5c(iv), $r$ was higher upon crystallization due to weaker coupling of the waveguide pair. These observations were based on the odd and even mode values from the simulated ring-bus coupling regions. The values can be found in the main supporting information document, Figure S4 and Table S7. Our devices started off by being over-coupled and moved towards critical-coupling or under-coupling conditions. This observation was supported by the fact that crystalline PCMs have a higher optical loss than its amorphous state, which increase the cavity roundtrip loss (or decrease in $a$). Given that the coupling strength is relatively unchanged, this leads to the shift of coupling condition from over-coupled ($r < a$) to under-coupled ($r > a$) [68]. In Figure 5c(ii), the device moved towards critical coupling when the PCM patch crystallizes as we see a narrower linewidth and sharper resonance contrast. However, in Figure 5c(iii), the device was slightly under-coupled as the loss in the crystalline state is greater by almost two-fold, based on Figure 5b. We observed that with a thicker PCM in Figure 5c(iv), the loss increased further and together with a further increase in $r$, the extent of under-coupling becomes larger. The FDE simulation of the capped 60 nm PCM single waveguide and its corresponding loss values can also be found in the main supporting information document, Figure S4c.

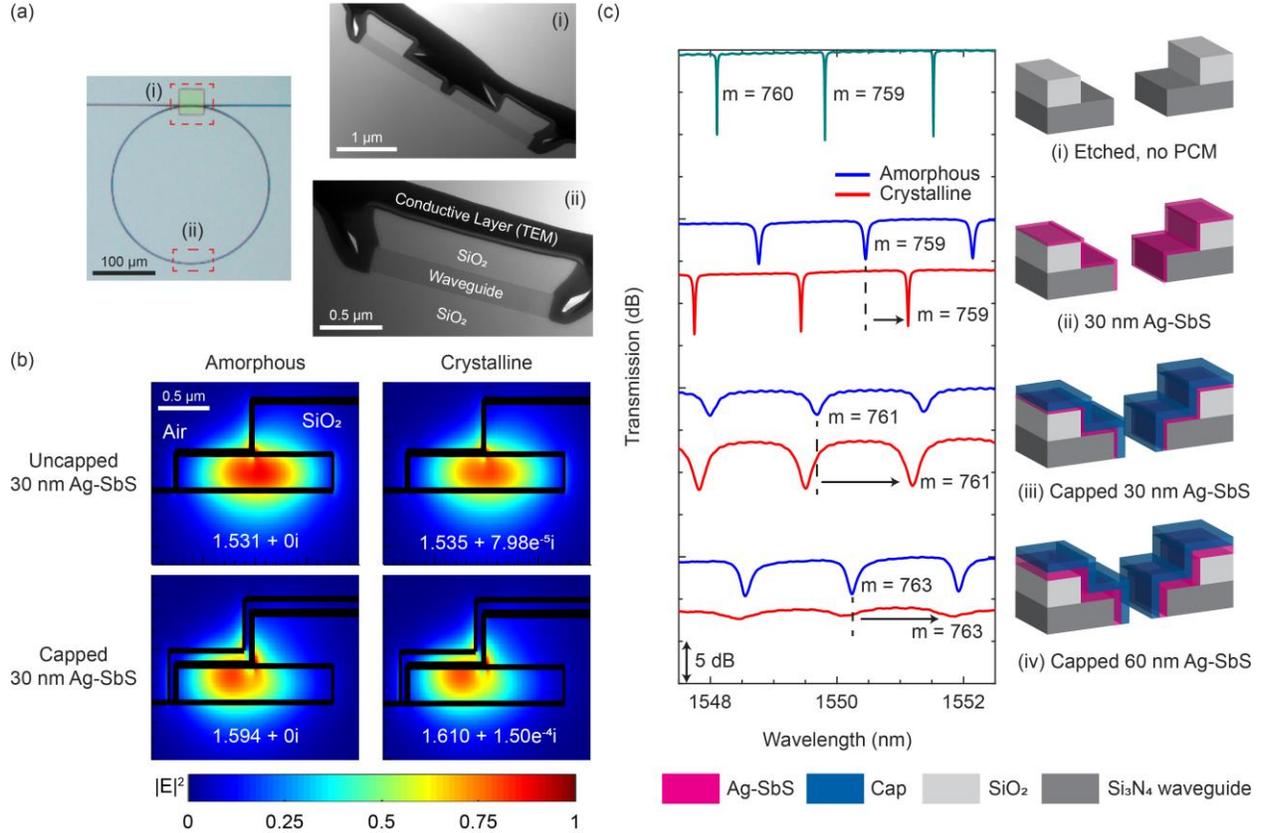

Figure 5: Comparison of uncapped and capped Ag-SbS MRR devices. (a) Microscope image of a MRR device with Ag-SbS patch deposited in the coupling region. (i) and (ii) show the TEM images of a separate MRR device after HF etching but before the PCM deposition step. Note, the TEM measurement locations correspond to the red boxes on the microscope image. Moreover, the device was coated with a conductive layer to address charging effects. (b) FDE simulation of PCM-deposited coupling regions of the MRR device (c) PCM-tuned MRR device spectra with various capping conditions, shown in (i) to (iv). The schematic illustrations of the devices are not drawn to scale.

Based on the ~$2\pi$ phase shift for the capped PCM with a fixed length, it is possible to reduce the cavity length by half, such that the free spectral range (FSR) doubles [68]. This would correspond to a phase shift of ~$\pi$, thereby changing the detuning from a resonance to an anti-resonance upon Ag-SbS structural phase transition. In addition, this $\pi$ phase shift maximizes extinction ratio, thereby improving switching resolution when multi-level switching functionalities are introduced [35]. Changing the cavity length will cause the Q-factor to decrease proportionally. To improve the Q-factor of smaller rings, the coupling gap between the ring and bus can be optimized to increase the cavity finesse [68]. Alternatively, the PCM strip length can be proportionally reduced to shorten the optical path length. Future works should investigate how the phase shift varies with various capped PCM strip lengths and cap thicknesses, similar to the experimental works in previous literature [32].

### 4) Conclusion

It is essential that capping layers are used to protect both $Sb_2S_3$ and Ag-SbS-tuned photonic devices. Moreover, the impact of capping layers must be considered when designing $Sb_2S_3$-based programmable devices. The caps clearly influence the device switching operations and optical performance. We recommend using at least a 30 nm thick ZnS-SiO$_2$ capping layer to protect Sb$_2$S$_3$ and Ag-SbS. Other caps [27, 32, 44, 45] could be considered, but both the interfacial energy and stress effects on the PCM structural phase transition needs to be carefully studied. Surprisingly, adding the lower refractive index ZnS-SiO2 cap to Ag-SbS increased the effect of Ag-SbS on the transmitted light. The effect was due to increased mode confinement in the PCM, leading to a greater change in $n_{eff}$ as the PCM crystallizes. This effect implies that capped couplers and MRRs can be designed to have a smaller footprint than their uncapped counterparts. This work shows that caps on PCMs are not just useful for stabilizing the PCM layer, but can also be used to tune the PCM crystallization temperature and reduce device footprint. Moreover, the capping layer can be exploited to enhance light-matter interactions with the PCM element. Future PCM-tuned photonic device designs should consider optimizing the capping layer to enhance device performance.

## 5) Author Contribution

T.Y.T, N. L., L.Y.T.L. and R.E.S. conceived the idea. All experiments and simulations were done by T.Y.T. after consulting: (a) R.E.S. and N. L. for the overall experimental design and execution and (b) A.S.K.T. and D.K.T.N. for top SiO$_2$ cladding etching. L.Y.M.T. helped with the MRR device calculation and analysis. Z.R. and C.L. assisted with the waveguide measurements. T.Y.T. wrote the manuscript with all authors contributing to it.

## 6) Funding

This work was funded by the Agency for Science, Technology and Research (A*STAR) grants A18A7b0058 and C220415015, and Ministry of Education (MOE) Singapore Tier 2 project (MOE-T2EP50220-0014).

## 7) Notes

The authors declare no conflict of interest.

## 8) Acknowledgements

We would like to thank Dr Yuyu Jiang (SUTD) for fruitful discussions, Mr Chern Chia Ser (SUTD) for conducting the HF etching training session, and Mr Zhonghua Gu (A*STAR IME) for the TEM measurements.

## 9) Supporting Information

The main supporting information document consists of the following: (1) EDX measurement of Ag-SbS film (2) Film thickness measurement method of the PCM crystallization thickness study in Figure 2 (3) Thickness measurements of ZnS-SiO$_2$ before and after annealing (4) Simulated reflectivity spectrum of uncapped and capped Sb$_2$S$_3$ and Ag-SbS (5) TEM cross section images of an etched MRR device (6) MRR azimuthal mode number derivation for devices in Figure 5c (7) Coupling condition analysis and FDE simulation of the various MRR devices in Figure 5c.

The second supporting information document consists of the raw data in Figure 4.

Videos 1 and 2 show the crystallization process of capped Sb$_2$S$_3$ and Ag-SbS respectively.

**Capping Layer Effects on Sb$_2$S$_3$-based Reconfigurable Photonic Devices: Supporting Information**

Ting Yu Teo[1,2]*, Nanxi Li[2]*, Landobasa Y. M. Tobing[2], Amy S. K. Tong[2], Doris K. T. Ng[2], Zhihao Ren[3], Chengkuo Lee[3], Lennon Y. T. Lee[2], and Robert Edward Simpson[4]*

[1] Singapore University of Technology and Design, 8 Somapah Road, Singapore 487372, Singapore
[2] Institute of Microelectronics (IME), Agency for Science, Technology and Research (A*STAR), 2 Fusionopolis Way, Innovis #08-02, Singapore 138634, Singapore
[3] Center for Intelligent Sensors and MEMS, Department of Electrical and Computer Engineering, National University of Singapore, 4 Engineering Drive 3, Singapore 117583, Singapore
[4] School of Engineering, University of Birmingham, Edgbaston, Birmingham B15 2TT, United Kingdom
Corresponding authors: tingyu_teo@mymail.sutd.edu.sg, linx1@ime.a-star.edu.sg, r.e.simpson.1@bham.ac.uk

**1) Energy Dispersive X-ray (EDX) Measurements of Ag-doped Sb$_2$S$_3$ films**

We measured the Ag-doped Sb$_2$S$_3$ thin film composition with EDX across 9 different regions on the sample wafer of dimension 1cm by 1cm. The measured compositions are shown in Table S1.

Table S1: EDX measurements of Ag-doped Sb$_2$S$_3$ film

| Region | Atomic Percentage | | |
|---|---|---|---|
| | Ag | Sb | S |
| 1 | 2.51 | 43.00 | 54.49 |
| 2 | 3.12 | 42.25 | 54.63 |
| 3 | 2.29 | 41.44 | 56.27 |
| 4 | 2.19 | 41.68 | 56.13 |
| 5 | 2.85 | 41.79 | 55.36 |
| 6 | 1.08 | 43.32 | 55.60 |
| 7 | 2.25 | 41.69 | 56.06 |
| 8 | 2.32 | 42.73 | 54.95 |
| 9 | 2.76 | 42.01 | 55.23 |
| Average | 2.37 | 42.21 | 55.41 |
| Standard deviation | 0.58 | 0.66 | 0.65 |

## 2) Atomic Force Microscope (AFM) Film Thickness Measurement

AFM was used to measure capped $Sb_2S_3$ film thicknesses before and after crystallization. Figures S1a and S1b show the camera image of a 55 nm-thick cap on $Sb_2S_3$ sample scan area before and after crystallization respectively. To extract the film thicknesses, z-height of the scanned areas were represented as histogram plots shown in Figures S1c and S1d. The distinct step on the sample caused the histogram to form two distinct peaks, which correspond to the film thickness.

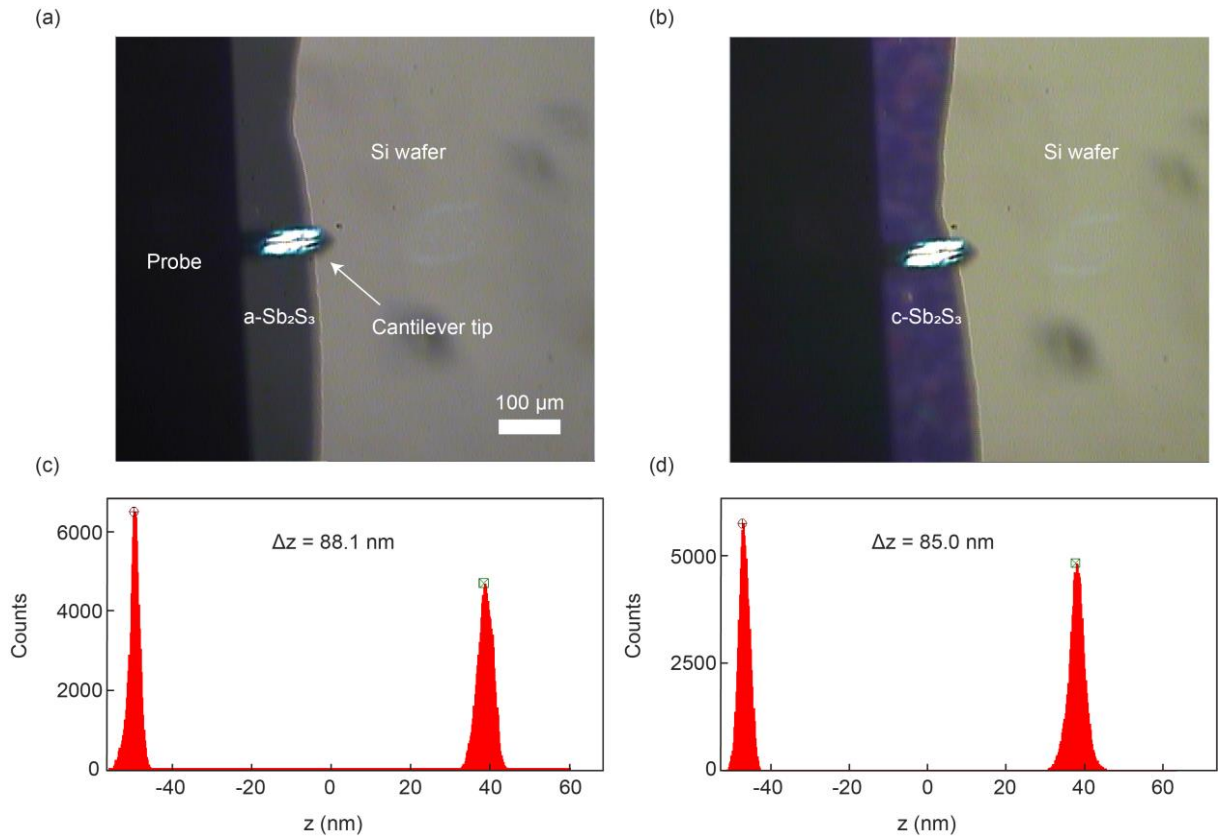

Figure S1: Camera image of the 55 nm-thick cap on $Sb_2S_3$ sample (a) before and (b) after crystallization. The corresponding (c) amorphous and (d) crystalline thickness values were derived through subtracting the z height values (x-axis) of the two distinct peaks in the histograms.

## 3) ZnS-SiO$_2$ Cap Thickness Before and After Thermal Annealing

The thickness of ZnS-SiO$_2$ films before and after thermal annealing are shown in Table S2. The measurement method was identical to the capped phase change material (PCM) thickness study in Figure 2. The single ZnS-SiO$_2$ films were annealed at 320 °C for 30 min. The heating parameters were chosen to match the conditions used to crystallize the $Sb_2S_3$ films.

Table S2: Thickness change of ZnS-SiO$_2$ thin films before and after thermal annealing

| Before Annealing (nm) | After Annealing (nm) |
|---|---|
| 39.9±0.3 | 40.5±0.3 |
| 70.1±0.3 | 70.2±0.2 |

## 4) Reflectivity Spectrum of PCM Thin Films

Through solving Fresnel's equation with the transfer matrix method on OpenFilters [1], we obtained the reflectivity spectrum of the film stacks shown in Figure S2. In the simulation model, optical constants of the PCM and ZnS-SiO$_2$ cap were from ellipsometry measurements while the Si substrate was from the OpenFilters built-in library. The gray region in the graphs represents the light source wavelength range. The reflectance readings of the capped crystalline films were generally lower than capped amorphous films in the light source wavelength range. Hence in Figures 3b and 3d, the film reflectance decreased upon crystallization.

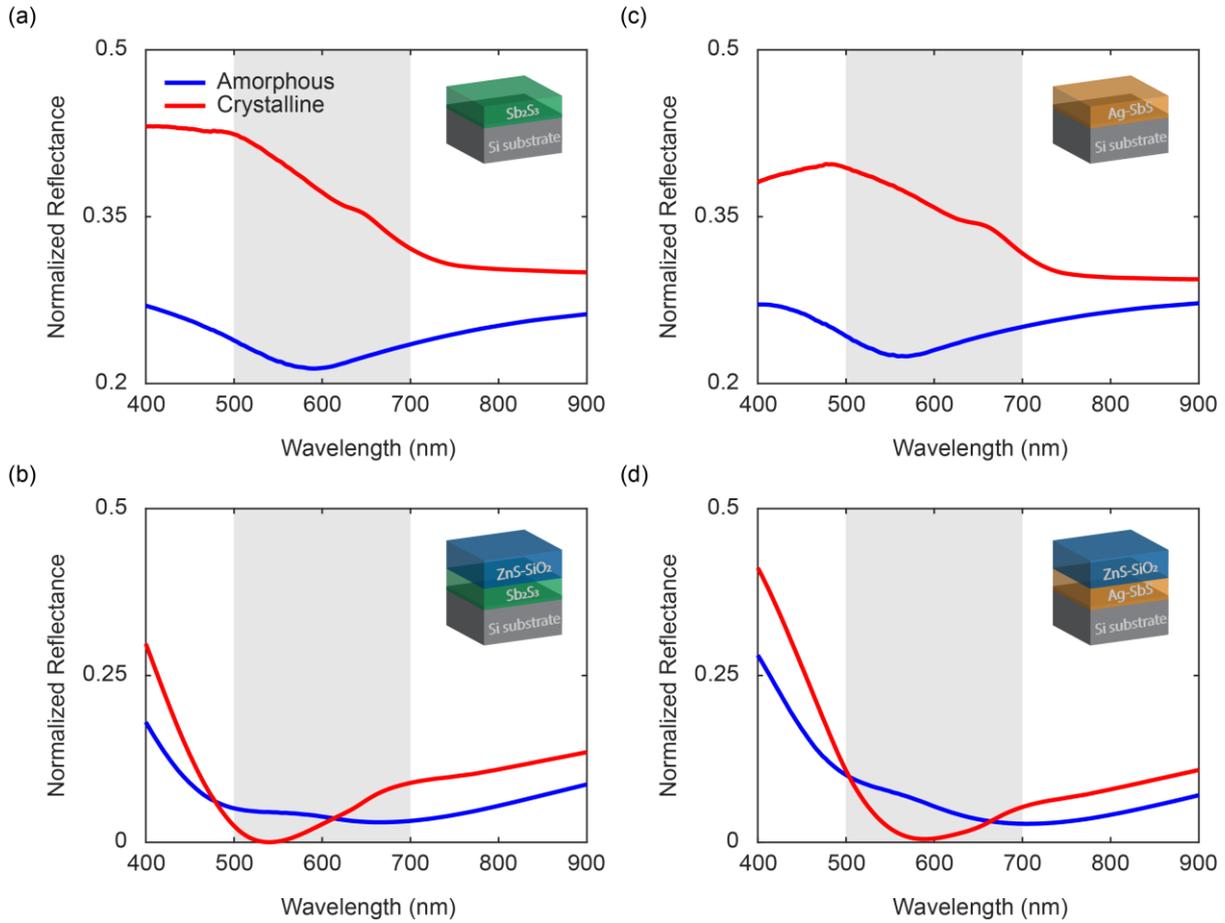

Figure S2: Simulated reflectance spectrum of the (a) uncapped and (b) capped Sb$_2$S$_3$ films and (c) uncapped and (d) capped Ag-SbS films. Figure insets show the corresponding simulation structure.

## 5) Transmission Electron Microscopy (TEM) Cross Section Image of Microring Resonator (MRR) Device

To verify that the HF etching process gives a uniform surface, TEM images were taken at different regions of the etched device as shown in Figure S3a. We observed that the top cladding oxide had consistent thickness throughout, except at the coupling region (Figure S3a(iii)) where the waveguides were exposed for PCM deposition. Figure S3b shows the cross-section of the pristine device.

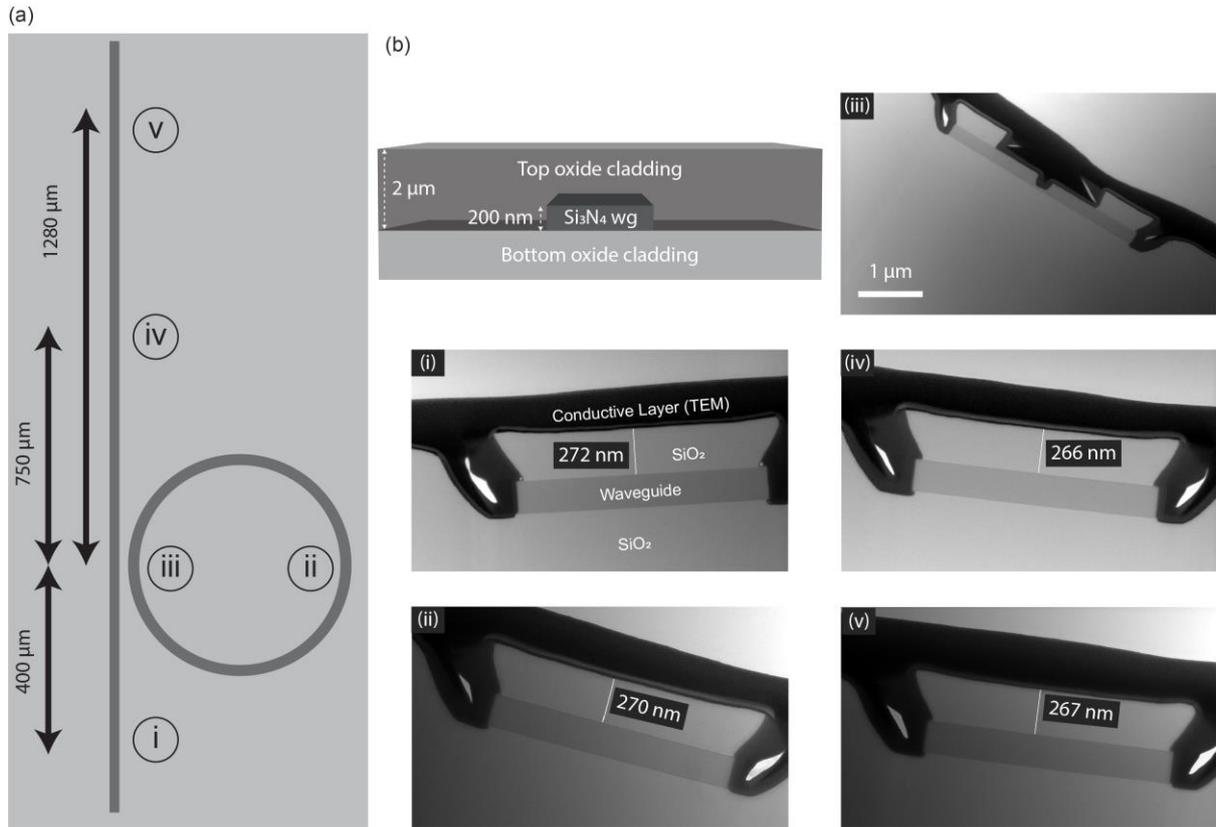

Figure S3: MRR device structure. (a) Top view schematic of the etched device. Labelled regions correspond to the TEM cross section measurements shown in (i) to (v). (b) Pristine device cross section before HF etching.

## 6) $m$ Value Approximations

The azimuthal mode number, $m$, is related to the optical path length by eq 1 [2]:

$$m\lambda_{res} = n_{eff} L_{cav} \tag{1}$$

where $\lambda_{res}$ is the resonant wavelength, $n_{eff}$ is the effective refractive index and $L_{cav}$ is the resonating cavity length.

For devices with uniform effective medium, $n_{eff}$ can be derived through an iterative calculation between the empirical group index, $n_g^{(exp)}$, and simulated group index, $n_g^{(sim)}$, as shown in Eqs. 2 and 3. Eq. 3 was derived from the dispersion relation between $n_{eff}$ and $n_g$ found in reference [2].

$$n_g^{(exp)} = \frac{\lambda_{res}^2}{FSR \times L_{cav}}, \tag{2}$$

where $FSR$ represents the free spectral range.

$$n_{eff} = A\lambda + n_g^{(sim)}(\lambda_{res}) \tag{3}$$

The gradient, $A$, and intercept, $n_g^{(sim)}$, gives a linear relationship between $\lambda$ and $n_{eff}$. $A$ and $n_g^{(sim)}$ can be obtained by solving Maxwell's equation of the waveguide stack across a wavelength range with Lumerical Eigenmode solver. For small $\lambda$ approximations around $\lambda_{res}$, $n_g^{(sim)}$ can be corrected with $n_g^{(exp)}$ so that the resultant $n_{eff}$ value, $n_{eff}^{(corrected)}$, is closer to its true value. Hence $m$ is approximated as such:

$$m\lambda_{res} = n_{eff}^{(corrected)} L_{cav} \tag{4}$$

The $m$ value for the device without the PCM and capping layer (Figure 5ci) was approximated using eqs 2 – 4. Table S3 shows the $m$ value approximation near the 1550 nm wavelength. Note, the resonating cavity has a 125 $\mu$m radius. We observed that $n_g^{(exp)}$ and $n_g^{(sim)}$ were not far off from each other, based on Table S3.

Table S3: $m$ value approximation of device in Figure 5c(i)

| Parameter | $\lambda_{res}$ (nm) | $FSR$ (nm) | $n_g^{(exp)}$ | $n_g^{(sim)}$ | $A$ | $n_{eff}^{(corrected)}$ | $m$ |
|---|---|---|---|---|---|---|---|
| Value | 1549.725 | 1.715 | 1.783 | 1.776 | -0.1845 | 1.497 | 759 |

For PCM-deposited MRR, $m$ is related to the optical path length as shown [3]:

$$m\lambda_{res} = n_{eff}^{(waveguide)}(L_{cav} - L_{PCM}) + n_{eff}^{(PCM)}(L_{PCM}) \tag{5}$$

$m$ of Figures 5c(ii) to 5c(iv) were approximated using $A$ and $n_g^{(exp)}$ from Table S3 to derive $n_{eff}^{(waveguide)}$. Moreover, $n_{eff}^{(PCM)}$ values were obtained from the single waveguide mode simulation shown in Figure S4. Tables S4 to S6 show the $m$ value approximation of the three devices in Figures 5c(ii) to 5c(iv). Note, $L_{PCM}$ is 40 $\mu$m long. Moreover, $\lambda_{res}$ corresponds to the resonant peak nearest to 1550 nm. As the capped devices have a phase shift close to $2\pi$, the nearest 1550 nm crystalline resonant peaks in Tables S5 and S6 corresponds to the next mode number. Hence, we see a greater $m$ value than the amorphous peaks.

Table S4: $m$ value approximation of device in Figure 5c(ii)

| Parameter | $\lambda_{res}$ (nm) | $n_{eff}^{(waveguide)}$ | $n_{eff}^{(PCM)}$ | $m$ |
|---|---|---|---|---|
| Amorphous | 1550.453 | 1.497 | 1.531 | 759 |
| Crystalline | 1551.125 | 1.497 | 1.535 | 759 |

Table S5: $m$ value approximation of device in Figure 5c(iii)

| Parameter | $\lambda_{res}$ (nm) | $n_{eff}^{(waveguide)}$ | $n_{eff}^{(PCM)}$ | $m$ |
|---|---|---|---|---|
| Amorphous | 1549.69 | 1.497 | 1.594 | 761 |
| Crystalline | 1549.508 | 1.497 | 1.610 | 762 |

Table S6: $m$ value approximation of device in Figure 5c(iv)

| Parameter | $\lambda_{res}$ (nm) | $n_{eff}^{(waveguide)}$ | $n_{eff}^{(PCM)}$ | $m$ |
|---|---|---|---|---|
| Amorphous | 1550.2375 | 1.497 | 1.672 | 763 |
| Crystalline | 1550.14 | 1.497 | 1.715 | 764 |

## 7) 2D Mode Simulation of Coupling Region

The mode pattern and $n_{eff}$ values of the MRR coupling region for various devices were obtained by solving Maxwell's equation with Lumerical Eigenmode solver. Figure S4 shows the corresponding mode profile of the 3 devices, together with their $n_{eff}$ and loss values. The coupling strength is denoted by the difference between the even and odd modes, $n_{even} - n_{odd}$. A greater difference gives a stronger coupling strength and smaller self-coupling coefficient, $r$. Table S7 shows $n_{even} - n_{odd}$ for each device. We observed that the coupling strength remained consistent for both capped and uncapped 30 nm Ag-SbS whilst the coupling strength is weaker (greater $r$) for the capped 60 nm Ag-SbS upon crystallization. Hence, all devices will tend towards under-coupling conditions as they experienced a greater optical loss with the crystalline PCM layer.

Table S7: $n_{even} - n_{odd}$ for the various devices

| Device | $n_{even} - n_{odd}$ (amorphous) | $n_{even} - n_{odd}$ (crystalline) |
|---|---|---|
| (ii) uncapped 30 nm Ag-SbS | 0.015 | 0.016 |
| (iii) 60 nm ZnS-SiO$_2$ on 30 nm Ag-SbS | 0.040 | 0.041 |
| (iv) 60 nm ZnS-SiO$_2$ on 60 nm Ag-SbS | 0.085 | 0.078 |

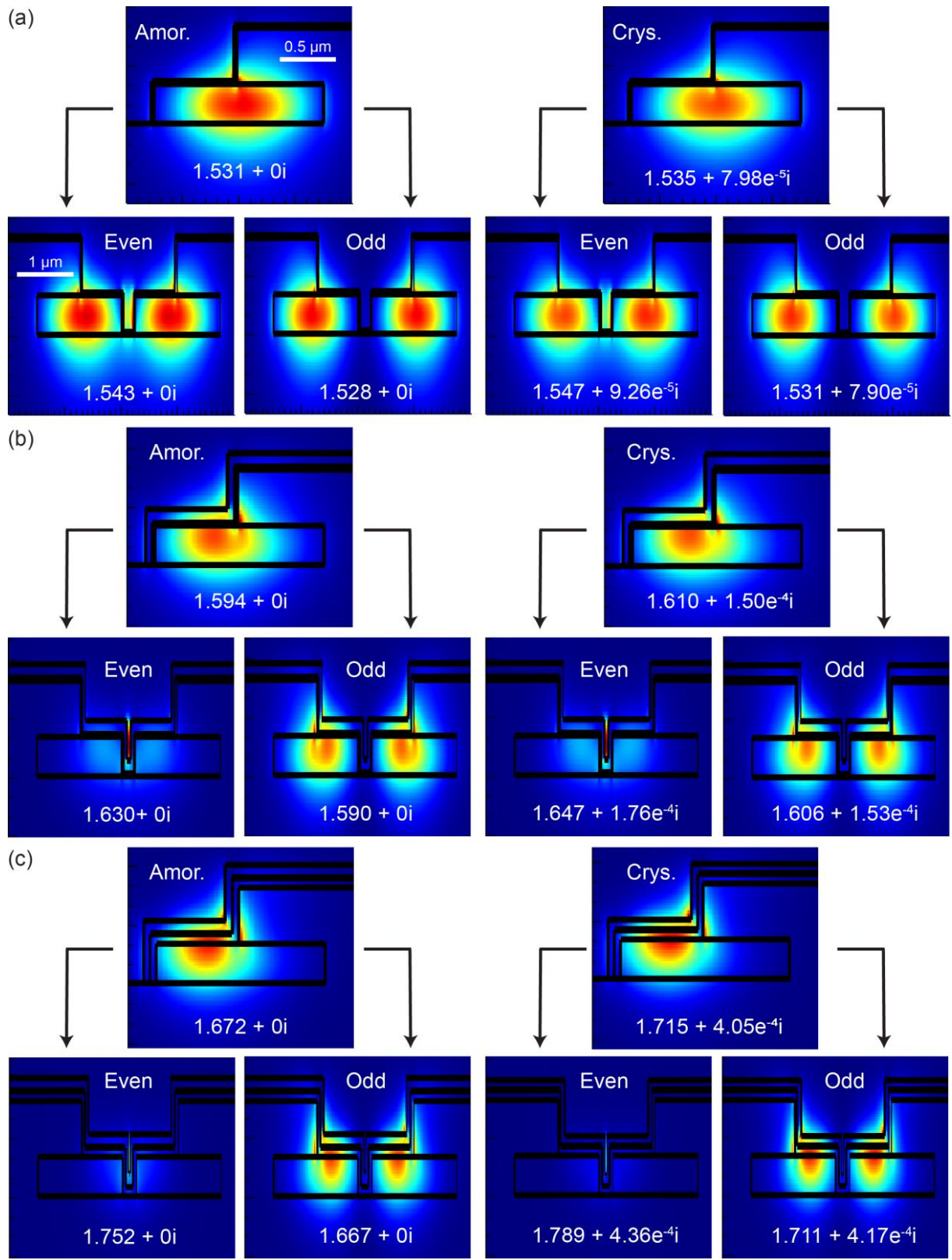

Figure S4: Simulated mode profile and corresponding $n_{eff}$ and loss values of (a) 30 nm Ag-SbS, (b) 60 nm ZnS-SiO$_2$ on 30 nm Ag-SbS and (c) 60 nm ZnS-SiO$_2$ on 60 nm Ag-SbS - tuned MRR devices at 1550 nm wavelength.